\documentclass[
 aip,
 jcp,
 amsmath,amssymb,
 reprint,
]{revtex4-1}

\usepackage{graphicx}
\usepackage{dcolumn}
\usepackage{bm}
\usepackage[utf8]{inputenc}
\usepackage[T1]{fontenc}
\usepackage{mathptmx}
\usepackage{etoolbox}
\usepackage{mathtools}
\usepackage{mathrsfs}
\usepackage{amsthm}
\usepackage{physics}
\usepackage{xcolor}
\usepackage{enumitem}
\usepackage{microtype}
\usepackage{hyperref}

\makeatletter
\def\@email#1#2{%
 \endgroup
 \patchcmd{\titleblock@produce}
  {\frontmatter@RRAPformat}
  {\frontmatter@RRAPformat{\produce@RRAP{*#1\href{mailto:#2}{#2}}}\frontmatter@RRAPformat}
  {}{}
}%
\makeatother

\newcommand{\R}{\mathbb{R}}

\newcommand{\FHK}{F_{\mathrm{HK}}}
\newcommand{\FLL}{F_{\mathrm{LL}}}
\newcommand{\Flieb}{F}
\newcommand{\Ts}{T_{\mathrm{s}}}
\newcommand{\EH}{E_{\mathrm H}}
\newcommand{\EHxc}{E_{\mathrm{Hxc}}}
\newcommand{\Exc}{E_{\mathrm{xc}}}
\newcommand{\vxc}{v_{\mathrm{xc}}}
\newcommand{\vH}{v_{\mathrm{H}}}
\newcommand{\vs}{v_{\mathrm{s}}}

\newcommand{\dom}{\operatorname{dom}}

\newcommand{\pair}[2]{\langle #1, #2\rangle}
\newcommand{\Cens}{\mathcal{C}^{\mathrm{ens}}}
\newcommand{\Cpure}{\mathcal{C}^{\mathrm{ps}}}
\newcommand{\Csens}{\mathcal{C}_{\mathrm s}^{\mathrm{ens}}}
\newcommand{\Cspure}{\mathcal{C}_{\mathrm s}^{\mathrm{ps}}}

\begin{document}

\preprint{AIP/123-QED}

\title[]
{Exact density-functional theory as parallel ensemble variational hierarchies: from Lieb's formulation to Kohn--Sham theory}

\author{Nan Sheng}
\affiliation{
Institute for Computational and Mathematical Engineering (ICME),
Stanford University, Stanford, CA 94305, USA.
}
\email{nansheng@stanford.edu}

\date{\today}

\begin{abstract}
Exact density-functional theory is recast here as two parallel exact
ensemble variational hierarchies: an interacting hierarchy rooted in
Lieb's ensemble formulation and a noninteracting hierarchy rooted in
exact noninteracting ensemble theory. In optimization terms,
\(N\)-representability is primal feasibility, Legendre--Fenchel duality
equates the primal and dual values, \(v\)-representability is dual
attainment, and the Hohenberg--Kohn theorem gives uniqueness, modulo
constants, of an attained local potential. The Kohn--Sham construction
couples the interacting density-space optimality condition to a
compatible noninteracting dual realization on their common
\(N\)-representable density domain.
State-class restrictions yield the Levy--Lieb and single-determinant
branches, while fractional particle number and fractional occupations
lead naturally to piecewise linearity, one-sided chemical potentials,
Janak-type relations, and the derivative discontinuity. This
organization locates the exactness of Kohn--Sham theory in the
preservation of the interacting density-space optimization together
with its compatible noninteracting realization, without implying a
general many-body spectral interpretation of Kohn--Sham eigenvalues.
\end{abstract}

\maketitle

\section{Introduction and historical perspective}

Ground-state density-functional theory is usually introduced through a
historical line beginning with the Hohenberg--Kohn theorem and leading to
the Kohn--Sham construction.\cite{HohenbergKohn1964,KohnSham1965}
This narrative is elegant and indispensable, but it compresses formally
distinct layers of the exact theory and can obscure useful logical
distinctions.

On the interacting side, the original Hohenberg--Kohn theorem, the Levy
constrained search, and Lieb's ensemble formulation are often presented as
successive variants of one basic idea.\cite{HohenbergKohn1964,Levy1979,
Lieb1983} Formally, however, they reorganize the theory in different ways.
The Hohenberg--Kohn theorem concerns ground-state
\(v\)-representable densities and the uniqueness, modulo constants, of
their supporting local potential. Ground-state nondegeneracy additionally
ensures a single-valued density-to-ground-state map, up to phase. The Levy
constrained search extends the functional description to the pure-state
\(N\)-representable density domain, while Lieb's formulation passes to
ensembles, places the theory in a Banach-space setting, and identifies the
closed convex density functional through Legendre--Fenchel duality with
the ground-state energy. These are closely related exact frameworks, but
they differ in admissible state class, supporting-potential domain, and
variational geometry.

A similar compression occurs on the noninteracting side. In many standard
presentations, noninteracting structure enters only when the Kohn--Sham
equations are introduced. Before constructing the Kohn--Sham auxiliary
description of the interacting problem, however, one may formulate the
exact noninteracting theory itself in ensemble variational form, with its
own source-side energy, fixed-density constrained-search functional,
Legendre--Fenchel duality, and representability structure.\cite{
Capelle2006,KvaalEkstromTealeHelgaker2014} The Kohn--Sham construction
in the stricter sense begins when the exact value-level decomposition is
combined with a compatible noninteracting supporting-potential
realization at the same density.

For the structural questions emphasized here, exact DFT is therefore
recast as two parallel exact ensemble variational hierarchies: an
interacting hierarchy and a noninteracting hierarchy, coupled by the
Kohn--Sham auxiliary construction. In the Coulomb setting considered
here, the two constrained-search theories share the same
\(N\)-representable density domain. Restricting the admissible state
class yields the Levy--Lieb pure-state branch on the interacting side and
the single-determinant branch on the noninteracting side. The latter gives
the familiar pure-state Kohn--Sham realization only when the density is
also noninteracting pure-state \(v\)-representable. Separately,
\(v\)-representability concerns the attainment of a local supporting
potential, while the Hohenberg--Kohn theorem concerns the uniqueness,
modulo constants, of such an attained interacting potential.

The same organization admits a direct optimization interpretation. The
fixed-density constrained search is the primal problem, and the scalar
potential is the dual variable associated with exact density
reproduction. In Lieb's ensemble setting, \(N\)-representability is
primal feasibility, the Legendre--Fenchel identity gives equality of the
primal and dual values, \(v\)-representability is dual attainment, and
the Hohenberg--Kohn statement gives uniqueness of the attained local
potential modulo constants. The exact noninteracting theory has the same
primal--dual structure, and Kohn--Sham theory couples its dual realization
to the interacting density-space optimality condition. Fractional
particle number and fractional occupations enter through the corresponding
ensemble extensions, while piecewise linearity, Janak-type relations, and
the derivative discontinuity reflect parametric and nonsmooth aspects of
the resulting variational structure.\cite{Lieb1983,Lammert2007,
PenzTellgrenCsirikRuggenthalerLaestadius2023PartI,Capelle2006,
KvaalEkstromTealeHelgaker2014}

This article offers a formal recasting of exact DFT that reorganizes
logical layers already present in the literature: the Hohenberg--Kohn and
Kohn--Sham papers,\cite{HohenbergKohn1964,KohnSham1965} the Levy and Lieb
formulations,\cite{Levy1979,Lieb1983} the exact-conditions literature on
fractional charge and the derivative discontinuity,\cite{
PerdewParrLevyBalduz1982,PerdewLevy1983,ShamSchluter1983} modern work on
representability and density-to-potential mappings,\cite{
Lammert2007,PenzTellgrenCsirikRuggenthalerLaestadius2023PartI} and ongoing
discussions of orbital energies and Janak-type interpretations.\cite{
Baerends2018,YangFan2024Orbital} The aim is not to replace the familiar
formulations, but to identify their precise roles within a common exact
ensemble and primal--dual variational structure.

The remainder of the paper is organized as follows. Section II develops
the interacting source-side energy, fixed-density constrained search, and
their Legendre--Fenchel relation, followed by fractional particle number,
the pure-state restriction, and Hohenberg--Kohn uniqueness on the
\(v\)-representable subdomain. Section III develops the parallel
noninteracting hierarchy, together with its supporting-potential,
orbital--occupation, and pure-state realizations. Section IV introduces
the exact Kohn--Sham density functional, couples the interacting and
noninteracting optimality conditions, and discusses the effective
potential, Janak relation, derivative discontinuity, gap decomposition,
and structural properties of the Hartree--exchange--correlation
remainder.

\section{Exact interacting DFT in the ensemble framework}

\subsection{Lieb's formulation, duality, and representability}

For the structural questions emphasized here, Lieb's density-matrix
formulation provides the natural broad starting point on the interacting
side. The state-to-density map is linear, the energy is linear in the
density matrix, and the admissible density-matrix set is convex. These
features place the exact theory in a setting adapted to convex duality,
subgradients, and supporting potentials.\cite{Lieb1983,Lammert2007}

We consider the standard nonrelativistic electronic Hamiltonian
\begin{equation}
H[v]=T+W+V[v].
\end{equation}
On an integer-\(M\) sector,
\begin{equation}
\begin{aligned}
T&=-\frac12\sum_{i=1}^{M}\Delta_i,
\quad
W=\sum_{1\le i<j\le M}\frac{1}{|\vb r_i-\vb r_j|},
\quad
V[v]=\sum_{i=1}^{M}v(\vb r_i).
\end{aligned}
\end{equation}
with the same symbols denoting the corresponding block-diagonal
Fock-space operators when variable particle number is allowed.
Throughout, \(M\in\mathbb N_0\) denotes an integer particle-number
sector, whereas \(N\) denotes a general average particle number and need
not be an integer.

The unrestricted Fock-space ground-state energy is
\begin{equation}
E[v]
=
\inf_{\Gamma\ge0,\ \tr\Gamma=1}
\tr\!\bigl[\Gamma\bigl(T+W+V[v]\bigr)\bigr].
\label{eq:Evdual}
\end{equation}
Here \(\Gamma\) ranges over positive trace-class density matrices on the
chosen electronic Fock space. Restricting \(\Gamma\) to the
\(M\)-electron sector gives \(E_M[v]\), whereas imposing
\(\tr(\Gamma\hat N)=N\) gives the fixed-average-particle-number value
\(E[v,N]\).

Passing from pure states to density matrices is an exact convex
relaxation at the potential level. Since
\(\Gamma\mapsto\tr[\Gamma H[v]]\) is linear, an ensemble energy is an
average of pure-state energies and cannot lie below the pure-state
minimum. The minimum is therefore unchanged, while mixtures within a
degenerate ground-state space are included naturally. For fixed
\(\Gamma\), the same expectation value is affine in \(v\); hence
\(E[v]\), as the pointwise infimum of affine functionals of \(v\), is
concave. We regard \(E\) as an extended-valued concave functional and
restrict ground-state statements to potentials for which the infimum is
finite and attained.

We write \(\Gamma\mapsto\rho\) when the density generated by \(\Gamma\),
\begin{equation}
\rho_\Gamma(\vb r)
=
\tr\!\left[\Gamma\hat n(\vb r)\right],
\end{equation}
equals the prescribed density \(\rho\). Grouping admissible states by
their densities defines
\begin{equation}
\Flieb[\rho]
=
\inf_{\Gamma\mapsto\rho}
\tr\!\bigl[\Gamma(T+W)\bigr],
\label{eq:LiebF}
\end{equation}
and gives
\begin{equation}
E[v]
=
\inf_{\rho}
\bigl(
\Flieb[\rho]+\pair{v}{\rho}
\bigr).
\label{eq:Liebdual2}
\end{equation}
The density constraint already fixes the average particle number through
\(\tr(\Gamma\hat N)=\int\rho\), so no separate particle-number
constraint is needed in the fixed-density search.

The density and potential spaces are taken as the dual pair
\begin{equation}
X=L^1(\R^3)\cap L^3(\R^3),
\qquad
X^\ast=L^\infty(\R^3)+L^{3/2}(\R^3),
\end{equation}
with
\begin{equation}
\pair{v}{\rho}
=
\int v(\vb r)\rho(\vb r)\,d\vb r.
\end{equation}
The space \(X\) is the ambient linear density space used for the convex
duality. On each fixed-\(M\) slice, the effective domain of \(\Flieb\)
is the usual finite-energy \(N\)-representable density class,
characterized by
\begin{equation}
\rho\ge0,
\qquad
\int\rho=M,
\qquad
\sqrt{\rho}\in H^1(\R^3).
\end{equation}
The Sobolev condition expresses finite-kinetic-energy representability
and, for the Coulomb Hamiltonian, permits an admissible representative
with finite interaction energy. The choice \(v\in X^\ast\) ensures that
the external-potential pairing is finite. We regard \(\Flieb\) as an
extended-valued functional on \(X\), with value \(+\infty\) outside its
effective domain.

Since \(\Flieb\) is proper, convex, and lower semicontinuous on \(X\),
the Fenchel--Moreau theorem gives\cite{Lieb1983}
\begin{equation}
\Flieb[\rho]
=
\sup_v
\bigl(
E[v]-\pair{v}{\rho}
\bigr).
\label{eq:Liebdual1}
\end{equation}
Together with Eq.~\eqref{eq:Liebdual2}, this is the
Legendre--Fenchel pair. It gives equality of the primal and dual values
but does not require the supremum to be attained. 

The same relation has a direct optimization interpretation. For fixed
\(\rho\), a trial dual variable \(v\) may be introduced through
\begin{equation}
\mathscr L(\Gamma,v;\rho)
=
\tr\!\bigl[\Gamma(T+W)\bigr]
+
\pair{v}{\rho_\Gamma-\rho}.
\label{eq:LiebLagrangian}
\end{equation}
Partial minimization over \(\Gamma\) gives the dual objective
\(E[v]-\pair{v}{\rho}\). Writing this Lagrangian does not assume that an
optimal multiplier exists. Only when the supremum in
Eq.~\eqref{eq:Liebdual1} is attained does the optimizing \(v\) become a
density-constraint multiplier and a supporting external potential for
\(\rho\).

Throughout, \(\partial G(x)\) denotes the global supporting
subdifferential: \(u\in\partial G(x)\) when
\begin{equation}
G(x')
\ge
G(x)+\pair{u}{x'-x}
\qquad
\text{for all admissible }x'.
\end{equation}
The superdifferential \(\partial^+E\) is defined by the reversed
supporting inequality for the concave functional \(E\). The
Legendre--Fenchel equality is attained precisely when
\begin{equation}
-v\in\partial\Flieb(\rho)
\qquad\Longleftrightarrow\qquad
\rho\in\partial^+E(v).
\label{eq:subgradrelation}
\end{equation}
Thus \(\rho\in\dom\Flieb\) expresses \(N\)-representability, whereas
\(\partial\Flieb(\rho)\neq\varnothing\) expresses attainment of a
supporting potential and hence interacting ensemble
\(v\)-representability.\cite{
Lammert2007,PenzTellgrenCsirikRuggenthalerLaestadius2023PartI}
The latter is generally a smaller, though dense, part of the admissible
density domain.~\cite{KvaalEkstromTealeHelgaker2014} At differentiable points the supporting relation reduces
to an Euler equation; at nondifferentiable points the subgradient form
remains exact.

The familiar Hohenberg--Kohn density-to-potential map therefore appears
only after dual attainment. It concerns the further uniqueness, modulo
constants, of an attained local supporting potential rather than the
existence of the universal functional on its full
\(N\)-representable domain.

\subsection{Fractional particle number and one-sided slopes}

Once the interacting theory is formulated in ensemble form, fractional
particle number enters naturally. One may impose particle number only in
expectation,
\begin{equation}
\tr \Gamma = 1,
\qquad
\tr(\Gamma \hat N)=N,
\end{equation}
with \(N\) not necessarily an integer.\cite{PerdewParrLevyBalduz1982}
This remains a zero-temperature exact variational problem and should not be
confused with thermal smearing. In this sense, fractional particle number
is not an external correction appended to exact DFT; it is already present
once the exact state space has been convexified.

Let \(E_M[v]\) denote the ground-state energy obtained by restricting the
search to the integer \(M\)-particle sector, and let \(E[v,N]\) denote the
minimum at fixed average particle number \(N\). Write
\(N=M+\omega\), where \(M=\lfloor N\rfloor\) and \(0\le\omega<1\). If
\(\Gamma_M\) and \(\Gamma_{M+1}\) are minimizing ensembles in the adjacent
integer sectors, then
\begin{equation}
\Gamma_N
=
(1-\omega)\Gamma_M+\omega\Gamma_{M+1}
\end{equation}
is admissible at average particle number \(N\). Since the expectation value
of the Hamiltonian is affine in the density matrix,
\begin{equation}
\tr\!\bigl[\Gamma_N H[v]\bigr]
=
(1-\omega)E_M[v]+\omega E_{M+1}[v].
\end{equation}
The ensemble construction therefore always provides an affine upper bound
between adjacent integer sectors. Under the usual convexity or
no-phase-separation assumption, that bound is exact and one obtains the
Perdew--Parr--Levy--Balduz (PPLB) piecewise-linearity
relation\cite{PerdewParrLevyBalduz1982}
\begin{equation}
E[v,N]
=
(1-\omega)E_M[v]+\omega E_{M+1}[v].
\label{eq:PPLBline}
\end{equation}
At \(\omega=0\), Eq.~\eqref{eq:PPLBline} reduces to
\(E[v,M]=E_M[v]\); cross-sector mixing is unnecessary, although mixtures
within a degenerate \(M\)-particle ground space may still be needed to
represent the density.
From the optimization viewpoint, fixing the average particle number is the
convex relaxation of the discrete sector problem: sector weights are
mixing variables and the Hamiltonian objective is affine. Under the same
convexity or no-phase-separation assumption, the adjacent integer
optimizers determine the relaxed optimum. Therefore the
fixed-average-particle-number ground-state branch is determined by the
integer-sector ground states. In
particular, the unrestricted ground-state energy is
\begin{equation}
E[v]=\inf_{M\in\mathbb N_0}E_M[v].
\label{eq:unrestricted-from-sectors}
\end{equation}
Thus, for the ground-state discussion below, the integer-\(M\)-sector
problems may be taken as the basic cases, while noninteger \(N\) is recovered
by the adjacent-sector PPLB extension. This reduction concerns the minimizing
ground-state ensembles and densities for a fixed external potential; the
universal constrained-search functional remains defined on its full
admissible density domain.

Once piecewise linearity holds, the relation to ionization energy,
electron affinity, and chemical potential becomes immediate. The left and
right derivatives at integer \(M\) are
\begin{equation}
\mu_M^-=\left.\dv{E[v,N]}{N}\right|_{M^-},
\qquad
\mu_M^+=\left.\dv{E[v,N]}{N}\right|_{M^+}.
\end{equation}
Because the energy is affine on each adjacent interval, these one-sided
derivatives are exactly the corresponding finite differences:
\begin{equation}
\begin{split}
\mu_M^- &= E[v,M]-E[v,M-1] = -I_M,
\\
\mu_M^+ &= E[v,M+1]-E[v,M] = -A_M.
\end{split}
\end{equation}
Thus ionization energy and electron affinity arise internally in exact DFT
as one-sided slope data of the interacting ensemble energy. For the
fixed-\(M\) constrained problem, the Lagrange multiplier associated with
the average-particle-number constraint therefore belongs to
\begin{equation}
\mu_M\in\partial_N E[v,M]
=
[\mu_M^-,\mu_M^+]
=
[-I_M,-A_M].
\end{equation}
Under \(v\mapsto v+C\), this interval shifts by \(C\), while the minimizing
\(M\)-particle states and densities remain unchanged and
\(E_M[v]\mapsto E_M[v]+CM\).

The same structure may be read in three equivalent languages. In energetic
language, one has the finite differences \(I\) and \(A\). In thermodynamic
language, one has the one-sided chemical potentials \(\mu_M^\pm\). In
convex-analytic language, one has the left and right supporting slopes of
the exact energy with respect to particle number. These are not distinct
facts but distinct descriptions of the same geometric structure. In the
same language, the familiar inequality \(I_M \ge A_M\) is equivalent to
\(\mu_M^- \le \mu_M^+\), that is, to a nonnegative jump between the
one-sided slopes at the integer.

This is why curvature between integers in approximate functionals is not
merely a numerical imperfection but a structural failure: it signals that
the approximation has lost the affine geometry implied by the exact
ensemble theory and therefore corrupts the exact slope data from which
electron addition and removal energetics are read.\cite{
CohenMoriSanchezYang2008Science,MoriSanchezCohenYang2008PRL,
HaitHeadGordon2018}

The many-body fundamental gap is therefore
\begin{equation}
E_{\mathrm g}^{\mathrm{true}}
=
I_M-A_M
=
\mu_M^+-\mu_M^-.
\end{equation}

\subsection{State-class and source-support structure:
Lieb, Levy--Lieb, and Hohenberg--Kohn}

The interacting theory contains two distinct but related restrictions.
The first concerns the state class used to represent a density: ensembles
versus pure states. The second concerns whether a density is merely
representable by such states, or is generated as a ground-state density
of some external potential.

For a fixed density, define the ensemble fiber
\begin{equation}
\Cens(\rho)
=
\qty{
\Gamma
\mid
\Gamma\ge0,\ \tr\Gamma=1,\ \Gamma\mapsto\rho
}.
\label{eq:Censfiber}
\end{equation}
The fiber is understood relative to the state space and
particle-number constraint chosen for the variational problem. Thus
\(\rho\) is \(N\)-representable precisely when
\(\Cens(\rho)\neq\varnothing\), and the Lieb functional is the
corresponding ensemble constrained search.

To compare the ensemble and pure-state branches, consider first a fixed
integer-\(M\) sector and define
\begin{equation}
\Cpure(\rho)
=
\left\{
\begin{array}{l}
\Gamma\in\Cens(\rho)\ \big|\ \operatorname{rank}\Gamma=1,\\
\Gamma\ \text{belongs to the }M\text{-electron sector}
\end{array}
\right\}.
\end{equation}
Then \(\rho\) is pure-state-representable precisely when
\(\Cpure(\rho)\neq\varnothing\). For the usual fixed-\(M\) finite-energy
density class, Harriman-type constructions show that every
\(N\)-representable density is also Slater representable.\cite{Harriman1981}
Thus the ensemble and pure-state searches share the same density domain;
they differ in the admissible states within each fiber and potentially in
the constrained-search value. 

The Levy--Lieb functional is the corresponding pure-state constrained
search,
\begin{equation}
\FLL[\rho]
=
\inf_{\Gamma\in\Cpure(\rho)}
\tr\!\bigl[\Gamma(T+W)\bigr]
=
\inf_{\Psi\mapsto\rho}
\mel{\Psi}{T+W}{\Psi}.
\label{eq:LLfunctional}
\end{equation}
Thus Levy--Lieb is obtained by narrowing the fixed-density fiber while
leaving the operator \(T+W\) unchanged. Since
\begin{equation}
\Cpure(\rho)\subseteq\Cens(\rho),
\end{equation}
one has
\begin{equation}
\Flieb[\rho]\le\FLL[\rho],
\end{equation}
with equality whenever an ensemble minimizer can be chosen pure.

We now turn to the separate sharper source-support restriction. A density is
interacting ensemble \(v\)-representable when it is generated by a
ground-state ensemble of some local scalar potential. In the global Lieb
geometry this is expressed by
\begin{equation}
\rho\in\dom\partial\Flieb,
\qquad
-v\in\partial\Flieb(\rho),
\end{equation}
while on a fixed-\(M\) density slice the supporting relation is understood
up to the constant particle-number multiplier. Equivalently, some
\(\Gamma\in\Cens(\rho)\) is a ground-state ensemble of
\(H[v]=T+W+V[v]\).

The pure-state \(v\)-representable class is defined analogously by
\begin{equation}
\rho\in\dom\partial\FLL,
\qquad
-v\in\partial\FLL(\rho),
\end{equation}
again up to an additive constant in the fixed-\(M\) setting. In
state-space language, this means that some pure state
\(\Psi\mapsto\rho\) satisfies
\begin{equation}
\Psi
\in
\operatorname*{argmin}_{\Psi'}
\mel{\Psi'}{H[v]}{\Psi'}.
\end{equation}
This is the traditional pure-state Hohenberg--Kohn setting.

We first establish potential uniqueness on the broader fixed-\(M\)
ensemble \(v\)-representable class. Ensembles may mix states within the
\(M\)-electron sector, including states in a degenerate ground-state
space, but do not yet mix different particle-number sectors.

Suppose that two potentials \(v\) and \(v'\) have ground-state ensembles
in the same fiber \(\Cens(\rho)\). The usual cross-variational argument
shows that each ensemble is also minimizing for the other potential.
Their supports therefore lie in a common ground-state subspace. For any
nonzero common ground state \(\Psi\),
\begin{equation}
H[v]\Psi=E_M[v]\Psi,
\qquad
H[v']\Psi=E_M[v']\Psi.
\end{equation}
Subtracting the two Schrödinger equations gives
\begin{equation}
\left[
\sum_{i=1}^{M}
\bigl(
v(\vb r_i)-v'(\vb r_i)
\bigr)
\right]\Psi
=
\bigl(E_M[v]-E_M[v']\bigr)\Psi.
\end{equation}
Under the standard regularity and unique-continuation assumptions, the
local multiplicative and particle-separable form of the source difference
then implies
\begin{equation}
v(\vb r)-v'(\vb r)=\mathrm{const.}
\end{equation}
almost everywhere.~\cite{PenzTellgrenCsirikRuggenthalerLaestadius2023PartI} Thus, on the fixed-\(M\) interacting ensemble
\(v\)-representable class,
\begin{equation}
\rho\longmapsto[v],
\end{equation}
where \([v]\) denotes the additive-constant equivalence class. A boundary
or gauge convention may select a representative of \([v]\).

The pure-state branch is the rank-one specialization of the same
uniqueness result. Nondegeneracy is not required for potential
uniqueness; it additionally ensures a single-valued
density-to-ground-state map, up to phase. In degenerate cases, the
density determines the potential class but need not determine a unique
ground-state wavefunction.

On the pure-state \(v\)-representable class, one may write
\begin{equation}
\FHK[\rho]
=
\mel{\Psi_\rho}{T+W}{\Psi_\rho},
\end{equation}
where \(\Psi_\rho\) is a chosen ground-state representative yielding
\(\rho\). On this class,
\begin{equation}
\FHK[\rho]=\FLL[\rho]=\Flieb[\rho].
\end{equation}
This equality does not follow from equality of the search spaces. Rather, since \(\pair{v}{\rho}\) is constant on the fiber, minimizing \(T+W\) there is equivalent to minimizing \(H[v]\); a ground
state yielding \(\rho\) does so already over all states. All three infima
are thus attained by the same ground state.

When the common density also admits global supporting potentials, the
same structure has a complementary interpretation in the Lieb-duality
geometry. If
\(-v,-v'\in\partial\Flieb(\rho)\) and
\begin{equation}
v_t=(1-t)v+t v',
\qquad
0\le t\le1,
\end{equation}
then
\begin{equation}
E[v_t]=(1-t)E[v]+tE[v'].
\end{equation}
Thus \(E\) is affine along the segment joining the two supporting
potentials. This is the dual-geometric signature of a common minimizing
face, but it does not by itself prove that the potentials differ only by
a constant. That conclusion still requires the local
Schrödinger-equation argument above.

The fixed-sector result extends to prescribed noninteger average particle
number. Suppose that two potentials possess constrained minimizing
ensembles with the same density \(\rho\). The crossed variational
inequalities again saturate, so a common constrained minimizer
\(\Gamma\) may be chosen. Since the Hamiltonian, particle-number
operator, and density operator preserve particle number, \(\Gamma\) may
be dephased with respect to the particle-number decomposition without
changing its density, energy, or average particle number.

Every occupied integer-particle-number block must then be a ground-state
ensemble of the corresponding fixed-sector Hamiltonian for both
potentials; otherwise that block could be lowered without changing its
weight or particle number. For \(N>0\), any occupied nonvacuum block
therefore reduces the problem to the fixed-sector result and gives
\begin{equation}
v(\vb r)-v'(\vb r)=\mathrm{const.}
\end{equation}
almost everywhere. Moreover, for every ensemble satisfying the
fixed-average-particle-number constraint,
\begin{equation}
\tr\!\left(\Gamma H[v+c]\right)
=
\tr\!\left(\Gamma H[v]\right)+cN.
\end{equation}
A constant shift changes every admissible energy by the same amount and
leaves the constrained minimizers unchanged. Hence
\(\rho\mapsto[v]\) also holds on the fixed-average-particle-number
ensemble \(v\)-representable class.

This statement should be distinguished from unrestricted Fock-space
minimization. Within each fixed integer sector,
\begin{equation}
E_M[v+c]=E_M[v]+cM,
\end{equation}
so different sectors undergo different energy shifts. A constant may
therefore change which sector contains the unrestricted ground state.
The separate sectorwise Hohenberg--Kohn relations cannot be combined
into a single unrestricted density-to-potential map modulo arbitrary
constants without an additional particle-number or chemical-potential
convention. In a grand-canonical formulation, the corresponding
invariance instead requires a joint shift of the scalar potential and
the chemical potential.

The interacting hierarchy should therefore be read along two independent
axes. Restricting \(\Cens(\rho)\) to \(\Cpure(\rho)\) gives the
Levy--Lieb pure-state branch. Separately, passing from
\(N\)-representability to \(v\)-representability requires attainment of
a supporting local potential, and the Hohenberg--Kohn statement supplies
its uniqueness modulo the fixed-\(M\) constant gauge. The traditional
pure-state Hohenberg--Kohn setting is obtained where these two
restrictions meet.

\section{Exact noninteracting DFT in the ensemble framework}

\subsection{The noninteracting constrained-search problem and its dual energy functional}

Throughout this section, the noninteracting ensemble theory refers to the
exact \(W=0\) density-functional theory associated with a one-body
Hamiltonian, formulated in ensemble variational form. This theory has its
own source-side energy, fixed-density constrained search, and
representability structure.

The unrestricted potential-side noninteracting ground-state energy is
\begin{equation}
E_{\mathrm s}[\vs]
=
\inf_{\Gamma \ge 0,\ \tr\Gamma =1}
\tr\!\bigl[\Gamma (T+V[\vs])\bigr].
\label{eq:Esdef}
\end{equation}
Here \(\Gamma\) ranges over positive trace-class density matrices on the
full electronic Fock space, exactly as on the interacting side.\cite{Capelle2006,
KvaalEkstromTealeHelgaker2014} Restricting the search to an integer sector
gives the corresponding sector energy, while imposing
\(\tr(\Gamma\hat N)=N\) gives \(E_{\mathrm s}[\vs,N]\). The subscript
\({\mathrm s}\) denotes the \(W=0\) objective and does not impose a
single-determinant state class.

As on the interacting side, passing from pure states to density matrices
is an exact convex relaxation at the potential level. Since the energy is
linear in \(\Gamma\), the minimum is unchanged, while degenerate mixtures
are included naturally. For fixed \(\Gamma\), the expectation value is
affine in \(\vs\); hence \(E_{\mathrm s}[\vs]\), as the pointwise
infimum of affine functionals of \(\vs\), is concave. We regard
\(E_{\mathrm s}\) as an extended-valued concave functional and restrict
ground-state statements to potentials for which the infimum is finite and
attained.

Grouping admissible states by their density defines the exact
noninteracting kinetic-energy functional
\begin{equation}
\Ts[\rho]
=
\inf_{\Gamma\mapsto\rho}
\tr(\Gamma T),
\label{eq:Tsdef}
\end{equation}
and gives
\begin{equation}
E_{\mathrm s}[\vs]
=
\inf_{\rho}
\bigl(
\Ts[\rho]+\pair{\vs}{\rho}
\bigr).
\label{eq:Tsdual2}
\end{equation}
As on the interacting side, the fixed-density constraint already fixes
\(N=\int\rho\). Thus \(\Ts\) is the noninteracting counterpart of
Lieb's fixed-density value function
In the Coulomb setting considered here, \(\Ts\) and \(\Flieb\) share
the same finite-energy \(N\)-representable density domain.\cite{Lieb1983}

On the same density--potential dual pair, \(\Ts\) is proper, convex, and
lower semicontinuous. Fenchel--Moreau duality therefore gives
\begin{equation}
\Ts[\rho]
=
\sup_{\vs}
\bigl(
E_{\mathrm s}[\vs]-\pair{\vs}{\rho}
\bigr).
\label{eq:Tsdual1}
\end{equation}
Together with Eq.~\eqref{eq:Tsdual2}, this is the noninteracting
Legendre--Fenchel pair. It is a value identity and does not require the
supremum to be attained. 

For a fixed target density, the same relation may be read through the
Lagrangian
\begin{equation}
\mathscr L_{\mathrm s}(\Gamma,\vs;\rho)
=
\tr(\Gamma T)
+
\pair{\vs}{\rho_\Gamma-\rho},
\end{equation}
where \(\vs\) is initially only a trial dual variable. Partial
minimization over \(\Gamma\) gives the dual objective
\(E_{\mathrm s}[\vs]-\pair{\vs}{\rho}\). Only when the supremum in
Eq.~\eqref{eq:Tsdual1} is attained does \(\vs\) become an optimal
density-constraint multiplier and a supporting noninteracting potential.
Equivalently,
\begin{equation}
-\vs\in \partial \Ts(\rho)
\qquad \Longleftrightarrow \qquad
\rho\in \partial^+ E_{\mathrm s}(\vs).
\label{eq:subgradTs}
\end{equation}
The functional \(\Ts\) may therefore be well defined even when no exact
supporting potential is attained. Noninteracting ensemble
\(v\)-representability is precisely this additional attainment property. However, the \(v\)-representable subdomains
\(\dom\partial\Ts\) and \(\dom\partial\Flieb\) are attainment sets of
two different dual problems and need not coincide.

Because both \(T\) and the density operator are one-body operators, the
value of the constrained search depends only on the one-particle density
matrix of \(\Gamma\). By the ensemble representability theorem for
fermionic 1RDMs, this 1RDM can be represented by a Slater-determinant
ensemble with the same one-body observables.~\cite{Coleman1963} At the level of infima, the
search in Eq.~\eqref{eq:Tsdef} may therefore be restricted to such
ensembles. If a minimizer exists, it may be chosen in that class. The
orbital--occupation representation follows from the \(W=0\) structure
rather than being imposed in Eq.~\eqref{eq:Tsdef}.

\subsection{Supporting potentials, state-level realization, and orbital representation}

Suppose now that \(\rho\) is noninteracting ensemble
\(v\)-representable, with \(N=\int\rho\). For integer \(N=M\), this is
the corresponding \(M\)-sector ground-state problem; noninteger \(N\) is
the adjacent-sector ensemble extension described above. In either case,
there exists a one-body potential \(\vs\) supporting \(\Ts\) at
\(\rho\) under the fixed-average-particle-number condition. The same
Hohenberg--Kohn-type uniqueness holds for the nonconstant part of this
\(W=0\) scalar-source problem.
Arbitrary constant shifts are gauge transformations on this slice, and a
supported density determines only the corresponding potential class,
\begin{equation}
\rho\longmapsto [\vs].
\end{equation}
Degeneracy may make the realizing orbital or ensemble representation
nonunique, without changing this supporting-potential structure.

Relative to the constrained-search problem
Eq.~\eqref{eq:Tsdef}, the attained supporting potential \(\vs\) is
the optimal dual multiplier associated with the density constraint.
When \(\Ts\) is differentiable at \(\rho\), this relation reduces to
the usual functional-derivative form, up to the fixed-\(N\) additive
constant.

For fixed \(\vs\), define the orbital--occupation energy
\begin{equation}
E_{\mathrm s}[\{n_i\},\{\phi_i\};\vs]
=
\sum_i n_i
\mel{\phi_i}{T+V[\vs]}{\phi_i}.
\label{eq:Esorbocc}
\end{equation}
From the fixed-density viewpoint, this is the state-dependent part of the
Lagrangian associated with the density constraint. Its minimum over
\(0\le n_i\le 1\), \(\sum_i n_i=N\), and orthonormal orbitals is the
potential-side value
\begin{equation}
E_{\mathrm s}[\vs,N]
=
\inf_{\{n_i\},\{\phi_i\}}
E_{\mathrm s}[\{n_i\},\{\phi_i\};\vs].
\label{eq:Esorbopt}
\end{equation}
When \(\vs\) supports \(\rho\), the corresponding Fenchel equality is
\begin{equation}
\Ts[\rho]
=
E_{\mathrm s}[\vs,N]-\pair{\vs}{\rho}.
\end{equation}
Orbital-rotation stationarity implies that the noninteracting 1RDM
commutes with \(T+V[\vs]\). Choosing a common eigenbasis gives the
one-body problem
\begin{equation}
\qty[-\frac12\Delta+\vs(\vb r)]\phi_i(\vb r)
=
\varepsilon_i\phi_i(\vb r),
\label{eq:onebodyvs}
\end{equation}
together with the occupation conditions stated below. The realizing
density is
\begin{equation}
\rho(\vb r)
=
\sum_i n_i |\phi_i(\vb r)|^2,
\qquad
0\le n_i\le 1,
\qquad
\sum_i n_i=N.
\label{eq:rhoocc}
\end{equation}
Here the fixed-average-particle-number condition appears explicitly as
\(\sum_i n_i=N\), completing the exact noninteracting ensemble
orbital representation of the density.\cite{
Capelle2006,KvaalEkstromTealeHelgaker2014}
For fixed \(\vs\), the one-body spectrum is fixed, so the remaining
occupation problem is linear. At stationary orbitals, the constrained
envelope theorem reduces the first variation to the explicit occupation
dependence,
\begin{equation}
\delta E_{\mathrm s}
=
\sum_i \varepsilon_i\,\delta n_i.
\end{equation}
The occupation KKT conditions therefore fill levels below a threshold
\(\mu_{\mathrm s}\), leave those above it empty, and permit fractional
occupation only at \(\varepsilon_i=\mu_{\mathrm s}\).

Consequently, \(E_{\mathrm s}[\vs,N]\) is convex and piecewise linear in
\(N\). On each open fractional-occupation segment,
\begin{equation}
\dv{E_{\mathrm s}[\vs,N]}{N}
=
\mu_{\mathrm s}
=
\varepsilon_{\mathrm f},
\label{eq:JanakNs}
\end{equation}
where \(\varepsilon_{\mathrm f}\) is the level being filled. At an
integer \(M\),
\begin{equation}
\partial_N E_{\mathrm s}[\vs,M]
=
[\varepsilon_{\mathrm H},\varepsilon_{\mathrm L}],
\end{equation}
so the HOMO and LUMO energies are the removal- and addition-side
endpoints. The kink has width
\(\varepsilon_{\mathrm L}-\varepsilon_{\mathrm H}\), the
noninteracting spectral gap
. 
This
fixed-potential piecewise linearity follows directly from the linear
occupation problem, unlike the interacting PPLB branch, where it rests
on the convexity assumption of Section~II.

Conversely, one may fix a target density \(\rho_{\mathrm{tar}}\) and ask
whether there exists a local noninteracting potential \(\vs\) whose
ground-state ensemble reproduces it. This is the inverse Kohn--Sham, or
inverse noninteracting DFT, problem. Existing inversion formulations may be
viewed as different practical and variational realizations of this
question.\cite{Zhao1994,WuYang2003,KhannaTribediKanungoGaviniZimmerman2026}
In optimization language, inverse Kohn--Sham theory seeks the attained dual
multiplier \(\vs\) associated with a prescribed primal density. A more
detailed variational interpretation is discussed in
Ref.~\onlinecite{sheng2026unified}.

This is also the point at which density reproduction and spectral
interpretation begin to diverge. The supporting potential \(\vs\) is
introduced to realize a given density within the exact auxiliary
framework. That does not, by itself, justify reading the resulting
one-particle spectrum as a literal many-body addition/removal spectrum of
the interacting problem. The exact role of the noninteracting framework is
therefore variational and density-reproductive first; any stronger
spectral interpretation belongs to a separate question and cannot be
assumed simply from the existence of a supporting potential.

\subsection{Pure-state specialization and noninteracting representability classes}

The common \(N\)-representable density domain has already been
established. What is specific to the \(W=0\) hierarchy is that both
\(T\) and \(\hat n(\vb r)\) are one-body operators, so the
constrained-search value depends only on the 1RDM. Coleman
representability therefore allows the search to be restricted, without
changing its infimum, to Slater-determinant ensembles.\cite{Coleman1963}

A Slater ensemble has the form
\begin{equation}
\Gamma_{\mathrm s}
=
\sum_I w_I\dyad{\Phi_I},
\qquad
w_I\ge 0,\qquad
\sum_I w_I=1,
\end{equation}
where each \(\Phi_I\) is a Slater determinant.  For a fixed density
\(\rho\), define the Slater-ensemble fiber
\begin{equation}
\Csens(\rho)
=
\qty{
\Gamma_{\mathrm s}
\mid
\Gamma_{\mathrm s}\text{ is a Slater ensemble},
\ \Gamma_{\mathrm s}\mapsto\rho
}.
\end{equation}
Then Eq.~\eqref{eq:Tsdef} may equivalently be written as
\begin{equation}
\Ts[\rho]
=
\inf_{\Gamma_{\mathrm s}\in \Csens(\rho)}
\tr(\Gamma_{\mathrm s}T).
\end{equation}
The corresponding single-determinant fiber is
\begin{equation}
\Cspure(\rho)
=
\qty{
\dyad{\Phi}
\mid
\Phi \text{ is a Slater determinant},
\ \Phi\mapsto\rho
}.
\end{equation}
The pure-state noninteracting constrained search is
\begin{equation}
\Ts^{\mathrm{ps}}[\rho]
=
\inf_{\Gamma_{\mathrm s}\in \Cspure(\rho)}
\tr(\Gamma_{\mathrm s}T).
\end{equation}
Thus \(\Ts^{\mathrm{ps}}\) is the single-determinant specialization of the
noninteracting constrained search.  Since
\(\Cspure(\rho)\subseteq\Csens(\rho)\),
\begin{equation}
\Ts[\rho]\le \Ts^{\mathrm{ps}}[\rho].
\label{eq:Ts-ineq}
\end{equation}
The difference between the two is therefore a genuine restriction of the
fixed-density search, not merely a different notation for the same object.

Noninteracting ensemble \(v\)-representability is the existence of a local
one-body potential \(\vs\) supporting \(\Ts\) at \(\rho\) on the
fixed-average-\(N\) slice, where \(N=\int\rho\), modulo an additive
constant. When the constrained-search infimum is attained, this is
equivalent to the existence of
\(\Gamma_{\mathrm s}\in\Csens(\rho)\) that is a ground-state Slater
ensemble of \(T+V[\vs]\).

The pure-state noninteracting version is the corresponding condition for
the single-determinant search: there exists a local potential \(\vs\),
defined modulo an additive constant on the fixed-\(N\) slice, such that
\begin{equation}
-\vs\in\partial\Ts^{\mathrm{ps}}(\rho).
\end{equation}
Equivalently, \(\rho\in\dom\partial\Ts^{\mathrm{ps}}\), with the
realizing ground state lying in \(\Cspure(\rho)\). Equality of the two constrained-search values, \(\Ts[\rho]=\Ts^{\mathrm{ps}}[\rho]\),
holds in particular when the minima are attained and a minimizing ensemble
can be chosen in the idempotent, or single-determinant, subclass.

At fixed density the relevant state fibers satisfy
\begin{equation}
\Csens(\rho)\subseteq \Cens(\rho),
\qquad
\Cpure(\rho)\subseteq \Cens(\rho),
\label{eq:statefiberchain}
\end{equation}
and
\begin{equation}
\Cspure(\rho)=\Csens(\rho)\cap\Cpure(\rho).
\end{equation}
These are inclusions of state fibers, not inclusions of interacting and
noninteracting \(v\)-representable density classes.  

At the level of the noninteracting one-body density matrix, the same
distinction appears as idempotent versus non-idempotent admissible
structure. The determinant restriction corresponds to idempotency, whereas
the broader ensemble formulation permits fractional occupations. Thus
fractional occupations and the distinction between ensemble and determinant
realizations belong to the exact noninteracting variational geometry.

\section{Exact Kohn--Sham theory as an auxiliary density-functional construction}

Sections II and III developed the interacting and noninteracting ensemble
theories separately. As established in Sec.~III, \(\dom\Flieb=\dom\Ts\); exact Kohn--Sham theory couples the two hierarchies on this common \(N\)-representable density domain.

\subsection{The Kohn--Sham auxiliary density functional}

The interacting density-space objective is
\begin{equation}
E_{\mathrm{KS}}[\rho]
=
\Flieb[\rho]+\pair{v}{\rho},
\qquad
\rho\in\dom\Flieb.
\label{eq:EKSprimary}
\end{equation}
Its unrestricted minimum is \(E[v]\), while imposing \(\int\rho=N\)
gives the fixed-average-particle-number value \(E[v,N]\).

The Kohn--Sham construction does not replace this interacting objective by
the bare noninteracting problem. It separates its exact noninteracting
kinetic contribution from the Hartree--exchange--correlation remainder
\(\EHxc[\rho]:=\Flieb[\rho]-\Ts[\rho]\). Thus
\begin{equation}
E_{\mathrm{KS}}[\rho]
=
\Ts[\rho]+\pair{v}{\rho}+\EHxc[\rho].
\label{eq:KSouter}
\end{equation}
This is the basic Kohn--Sham split between the noninteracting kinetic
energy and the remaining interaction contribution.

To resolve the structure of \(\EHxc\), let
\(\Gamma_{\mathrm s}\mapsto\rho\) be a minimizing ensemble for the
noninteracting constrained search, chosen, when the minimum is nonunique,
to minimize \(\tr(\Gamma W)\) within the minimizing face. The classical
part of the interaction is the Hartree energy
\begin{equation}
\EH[\rho]
=
\frac12
\iint
\frac{\rho(\vb r)\rho(\vb r')}{|\vb r-\vb r'|}
\,d\vb r\,d\vb r'.
\end{equation}
Exchange and correlation are defined by
\begin{equation}
E_{\mathrm x}[\rho]
=
\tr(\Gamma_{\mathrm s}W)-\EH[\rho],
\qquad
E_{\mathrm c}[\rho]
=
\Flieb[\rho]-\tr\!\bigl[\Gamma_{\mathrm s}(T+W)\bigr].
\end{equation}
Since \(\tr(\Gamma_{\mathrm s}T)=\Ts[\rho]\), these definitions give
\(\EHxc[\rho]=\EH[\rho]+\Exc[\rho]=\EH[\rho]+E_{\mathrm x}[\rho]+E_{\mathrm c}[\rho]\), with
\(\Exc:=E_{\mathrm x}+E_{\mathrm c}\). Exchange is the nonclassical
interaction contribution already present in the exact noninteracting
reference, whereas correlation is the remaining change produced by the
fully interacting optimization.

The same functional also has an equivalent orbital--occupation variational
form. Using the same symbol for the lifted objective, define
\begin{equation}
\begin{aligned}
E_{\mathrm{KS}}[\{n_i\},\{\phi_i\}]
={}&
\sum_i n_i\mel{\phi_i}{T}{\phi_i}
+\pair{v}{\rho}+\EHxc[\rho],
\\
\rho(\vb r)
={}&
\sum_i n_i|\phi_i(\vb r)|^2.
\end{aligned}
\label{eq:KSorbocc}
\end{equation}
For fixed \(\rho\), \(E_{\mathrm{KS}}[\rho]\) is the infimum of
Eq.~\eqref{eq:KSorbocc} over all orthonormal orbital--occupation
representations of that density with \(0\le n_i\le1\).
Consequently, minimizing \(E_{\mathrm{KS}}[\rho]\) over densities is
equivalent to a single minimization over the admissible orbitals and
occupations. This is an identity of infima; an arbitrary representation
of \(\rho\) need not attain \(\Ts[\rho]\).

The decomposition preserves the interacting density objective while
realizing its kinetic contribution through the noninteracting hierarchy.
In optimization terms, \(E_{\mathrm{KS}}\) is the exact density-space
primal objective, while the noninteracting problem supplies a dual
realization of the \(\Ts\) first-order term. Its exactness lies in
reproducing the same minimizing density and energy, not in identifying the
interacting and noninteracting theories. The identity above concerns
functional values on the common \(N\)-representable density domain and
does not by itself imply a decomposition of first-order information. A
local Kohn--Sham realization additionally requires compatible interacting
and noninteracting supporting structures at the same density.

Since \(\Flieb\) is convex and the external-potential term is linear,
\(E_{\mathrm{KS}}\) is convex. Its constrained subdifferential optimality
condition is therefore sufficient for global minimality, although the
minimizing density need not be unique. Direct density optimization is
possible in principle wherever suitable functional values or first-order
information are available. Practical Kohn--Sham theory instead approximates \(\Exc\) and
solves the coupled optimality system self-consistently through the
noninteracting problem.

\subsection{Stationarity, effective potentials, and Janak's relation}

Let \(\rho\) be a fixed-\(M\) minimizing density of
\(E_{\mathrm{KS}}\). Since
\(E_{\mathrm{KS}}[\rho]=\Flieb[\rho]+\pair{v}{\rho}\), its
minimizing property already supplies the interacting supporting
relation on the fixed-\(M\) density slice. A local noninteracting
realization requires the additional assumption that the same density is
noninteracting ensemble \(v\)-representable, so that the inner
fixed-density search admits an attained local supporting potential
\(\vs\).

When the relevant first-order derivatives exist, define
\begin{equation}
v_{\mathrm{Hxc}}(\vb r)
=
\frac{\delta E_{\mathrm{Hxc}}[\rho]}
{\delta\rho(\vb r)}
=
\vH(\vb r)+\vxc(\vb r).
\end{equation}
Matching the outer density optimality condition with the inner
noninteracting supporting relation gives
\begin{equation}
[\vs]=[v+v_{\mathrm{Hxc}}],
\end{equation}
where the additive constant is the multiplier associated with the
fixed-particle-number constraint. Choosing a common representative
yields
\begin{equation}
\vs(\vb r)
=
v(\vb r)+\vH(\vb r)+\vxc(\vb r).
\label{eq:vseff}
\end{equation}

For fixed occupations, orbital stationarity of the inner
noninteracting problem implies that its one-body density matrix
commutes with \(T+V[\vs]\). Choosing a common eigenbasis gives the
Kohn--Sham equations
\begin{equation}
\qty[-\frac12\Delta+\vs(\vb r)]\phi_i(\vb r)
=
\varepsilon_i\phi_i(\vb r),
\end{equation}
together with the density and occupation conditions in
Eq.~\eqref{eq:rhoocc}. The particle-number multiplier is the
corresponding occupation threshold \(\mu\).

For fixed occupations, let
\(E_{\mathrm{KS}}[\{n_i\}]\) denote the stationary total-energy value
obtained after the density and orbitals have been determined
self-consistently along a regular branch. The constrained envelope
theorem then gives
\begin{equation}
\frac{\partial E_{\mathrm{KS}}[\{n_j\}]}{\partial n_i}
=
\mel{\phi_i}{T+V[\vs]}{\phi_i}
=
\varepsilon_i.
\label{eq:Janak}
\end{equation}
This is Janak's relation for the full Kohn--Sham total energy.~\cite{Janak1978} The
density-dependent terms are not omitted from the derivative; their
first-order contribution is contained in \(\vs\) through
Eq.~\eqref{eq:vseff}.

Minimization over the occupations subject to
\(0\le n_i\le1\) and \(\sum_i n_i=N\) gives the same threshold
conditions as in the fixed-potential noninteracting problem. Along a
smooth fractional-occupation branch,
\begin{equation}
\dv{E[v,N]}{N}
=
\mu
=
\varepsilon_{\mathrm f},
\label{eq:JanakN}
\end{equation}
where \(\varepsilon_{\mathrm f}\) is the fractionally occupied
frontier level. Combined with the PPLB result of Section~II, this
implies that \(\varepsilon_{\mathrm f}\) is constant on each open
segment, taking the values \(-I_M\) and \(-A_M\) on the removal and
addition sides, respectively. Unlike the fixed-\(\vs\) result of
Section~III, the potential now changes self-consistently with \(N\);
its frontier eigenvalue must nevertheless remain constant. This
condition underlies analyses of delocalization error in approximate
functionals.\cite{
CohenMoriSanchezYang2008Science,MoriSanchezCohenYang2008PRL}

Equation~\eqref{eq:Janak} is an occupation-space stationarity
statement. It does not, by itself, assign a general many-body
ionization, affinity, or excitation-energy interpretation to every
Kohn--Sham eigenvalue.\cite{Baerends2018,YangFan2024Orbital}

At an integer \(M\), the one-sided form of
Eq.~\eqref{eq:JanakN} matches the interacting PPLB slopes. The
removal-side frontier level is the HOMO, giving
\(\varepsilon_{\mathrm H}=-I_M\). On the addition side, the
self-consistent limiting potential differs from the integer-\(M\)
Kohn--Sham potential by a spatial constant. Since \(v\) is fixed and
the Hartree potential has the same limiting density, this jump belongs
to \(\vxc\) and defines the exchange-correlation derivative
discontinuity \(\Delta_{\mathrm{xc}}\).~\cite{PerdewLevy1983,ShamSchluter1983} Hence
\begin{equation}
\partial_N E[v,M]
=
\bigl[
\varepsilon_{\mathrm H},
\varepsilon_{\mathrm L}+\Delta_{\mathrm{xc}}
\bigr]
=
[-I_M,-A_M],
\label{eq:KSmultiplierset}
\end{equation}
where \(\varepsilon_{\mathrm L}\) is the LUMO of the integer-\(M\)
Kohn--Sham potential. Equality of the interval widths gives
\begin{equation}
E_{\mathrm g}^{\mathrm{true}}
=
I_M-A_M
=
\bigl(
\varepsilon_{\mathrm L}-\varepsilon_{\mathrm H}
\bigr)
+
\Delta_{\mathrm{xc}}
\equiv
E_{\mathrm g}^{\mathrm{KS}}
+
\Delta_{\mathrm{xc}}.
\end{equation}
Thus the bare Kohn--Sham gap is the width of the fixed-potential
multiplier interval, while \(\Delta_{\mathrm{xc}}\) supplies the
addition-side shift required by the self-consistent physical branch.

\subsection{Structural properties of the exchange-correlation functional}

The basic Kohn--Sham remainder satisfies
\(\EHxc[\rho]=\EH[\rho]+\Exc[\rho]\). Since \(\EH\) is explicit, the nontrivial exact
structure resides in \(\Exc\), which combines the exchange contribution
already present in the exact noninteracting reference with the remaining
correlation contribution required to match the fully interacting
constrained search. It is therefore not merely a residual Coulombic
correction, but the structured interface quantity between two exact but
nonidentical variational theories.

One important family of exact properties concerns uniform coordinate
scaling.\cite{Levy1985Scaling,PerdewLevy1997,EngelDreizler2011}
If the density is scaled as
\begin{equation}
\rho_\gamma(\vb r)=\gamma^3\rho(\gamma\vb r),
\qquad \gamma>0,
\end{equation}
then the corresponding exact functionals obey
\begin{equation}
\Ts[\rho_\gamma]=\gamma^2\Ts[\rho],
\qquad
\EH[\rho_\gamma]=\gamma \EH[\rho].
\end{equation}
Exchange scales linearly,
\begin{equation}
E_{\mathrm x}[\rho_\gamma]=\gamma E_{\mathrm x}[\rho],
\end{equation}
whereas correlation obeys the more subtle coupling-constant-related
scaling structure familiar from exact DFT.\cite{Levy1985Scaling,
PerdewLevy1997}
These different scaling laws reflect the different roles of exchange and
correlation in comparing the interacting and noninteracting theories.

Another exact structure concerns the one-electron limit and
self-interaction cancellation. For every one-electron density, the exact
correlation energy vanishes and exchange exactly cancels the Hartree
self-repulsion:
\begin{equation}
E_{\mathrm c}[\rho_{1e}]=0,
\qquad
E_{\mathrm x}[\rho_{1e}]=-\EH[\rho_{1e}],
\qquad
\EHxc[\rho_{1e}]=0.
\end{equation}
This is the exact one-electron self-interaction
cancellation.\cite{PerdewZunger1981,KaplanLevyPerdew2023}
Its significance is both formal and practical. Formally, it shows that the
interfacial quantity \(\Exc\) is already highly constrained in the
simplest possible density sector. Practically, it explains why
self-interaction errors in approximate functionals are structural failures
rather than minor empirical defects.

Global energetic bounds provide another family of exact constraints. The
best known example is the Lieb--Oxford bound, which places a universal
lower bound on the exchange-correlation energy.\cite{LiebOxford1981,
Levy1993TightBound,OdashimaCapelle2007} In one common form,
\begin{equation}
\Exc[\rho]\ge -C_{\mathrm{LO}}\int \rho(\vb r)^{4/3}\,d\vb r,
\end{equation}
with a universal constant \(C_{\mathrm{LO}}\). The precise optimal value
of the constant is not needed here. What matters for the present
discussion is that the exact remainder term is subject to global many-body
restrictions independent of any specific approximation. Closely related
local structure may also be expressed through exchange-correlation hole
constraints.\cite{PerdewBurkeErnzerhof1996,KaplanLevyPerdew2023}

The adiabatic connection is especially important in the present
framework.\cite{Levy1996AdiabaticConnection,
Yang1998GeneralizedAdiabaticConnection,TealeCorianiHelgaker2010}
The coupling-constant-resolved constrained-search functional is
\begin{equation}
F^\lambda[\rho]
=
\inf_{\Gamma\mapsto\rho}
\tr\!\bigl[\Gamma(T+\lambda W)\bigr].
\end{equation}
For the Coulomb family considered here, the functionals \(F^\lambda\),
\(0\leq\lambda\leq1\), share the same \(N\)-representable density domain,
while their \(v\)-representable subsets may depend on \(\lambda\). Let
\(\Gamma_\rho^\lambda\) denote a selected minimizing state. Then
\begin{equation}
W_{\mathrm{xc}}^\lambda[\rho]
=
\tr\!\bigl[\Gamma_\rho^\lambda W\bigr]-\EH[\rho],
\end{equation}
and the exchange-correlation energy is
\begin{equation}
\Exc[\rho]
=
\int_0^1 W_{\mathrm{xc}}^\lambda[\rho] \, d\lambda.
\end{equation}
At fixed \(\rho\), \(F^\lambda[\rho]\) is a parametric primal value
function. The constrained envelope theorem gives, at the noninteracting
endpoint, the Hartree--exchange interaction of the selected minimizing
ensemble:
\begin{equation}
\left.\dv{F^\lambda[\rho]}{\lambda}\right|_{0^+}
=
\tr(\Gamma_{\mathrm s}W)
=
\EH[\rho]+E_{\mathrm x}[\rho].
\end{equation}
Thus \(\Exc\) is the density-fixed interaction-strength integral
connecting the noninteracting and interacting endpoints along a
density-preserving path.

The same family admits a parallel primal--dual interpretation. For each
fixed \(\lambda\), \(F^\lambda[\rho]\) is the fixed-density primal value
for the internal Hamiltonian \(T+\lambda W\), and the corresponding
potential-side ground-state energy is its dual value. A supporting
\(v^\lambda\), when it exists, is the attained dual optimizer. The
pure-state analogue is obtained by restricting the same fiber to pure
states, and the endpoint ensemble theories are
\begin{equation}
F^0[\rho]=\Ts[\rho],
\qquad
F^1[\rho]=\Flieb[\rho].
\end{equation}
The \(\lambda=1\) pure-state branch gives the Levy--Lieb functional. For
every fixed \(\lambda\), the same scalar-source variational structure also
carries a Hohenberg--Kohn-type uniqueness statement: on the corresponding
\(v\)-representable class,
\begin{equation}
-v^\lambda\in\partial F^\lambda(\rho)
\quad\Longrightarrow\quad
\rho\longmapsto [v^\lambda].
\end{equation}
Thus \(\lambda=0\) gives the noninteracting constrained search and the
uniqueness of the Kohn--Sham potential class, \(\lambda=1\) gives the
physical interacting Lieb/Hohenberg--Kohn structure, and intermediate
values give the partially interacting analogues. In this sense, the
adiabatic connection is the coupling-constant-resolved family linking the
noninteracting and interacting density-side hierarchies already isolated
in the present reconstruction. At fixed density \(\rho\), the map
\begin{equation}
\lambda\mapsto F^\lambda[\rho]
\end{equation}
is the infimum, over all admissible states yielding \(\rho\), of affine
functions \(\tr[\Gamma(T+\lambda W)]\) of \(\lambda\); since \(W\ge0\),
these have nonnegative slopes, so the infimum is concave and monotone
nondecreasing. The adiabatic-connection formula should thus not be read
as requiring naive everywhere differentiability in \(\lambda\); rather,
one-sided derivatives exist everywhere, with differentiability almost
everywhere the natural generic situation. By the envelope theorem, these
one-sided derivatives equal \(\tr[\Gamma_\lambda W]\) at a minimizer on
the corresponding side, independent of how \(\Gamma_\lambda\) varies with
\(\lambda\); at a kink the minimizing set is degenerate and this value
ranges over the two one-sided derivatives, but since kinks are countable
the ambiguity does not affect
\(F^1[\rho]-F^0[\rho]=\int_0^1 F'_+(\lambda)\,d\lambda\), which holds by
absolute continuity of the concave function.

A common density domain does not imply a common supporting-potential class.
A fully realized adiabatic-connection path requires the same density
\(\rho\) to admit a supporting \(v^\lambda\) at every
\(\lambda\in[0,1]\).

Seen from this angle, several exact properties line up naturally. Scaling
relations show that exchange and correlation transform differently because
they encode different parts of the interacting/noninteracting comparison.
The one-electron limit shows how self-interaction must cancel exactly. The
Lieb--Oxford bound and hole constraints express global and local
restrictions on the same interfacial quantity. The adiabatic connection
shows that \(\Exc\) is the density-fixed interaction-strength integral
linking the two endpoint theories, with potential-side realization
requiring a supporting \(v^\lambda\) at each intermediate coupling
strength. Taken together, these properties support the central claim of
this section: in exact DFT, the exchange-correlation functional is the
structured interface between two exact variational hierarchies.

This is also the methodological reason exact constraints matter for
approximation.\cite{PerdewRuzsinszkyTaoStaroverovScuseriaCsonka2005,
KaplanLevyPerdew2023}
Their importance does not come merely from their usefulness as design
principles for density-functional approximations. It comes from the fact
that they arise from the exact theory itself and therefore reflect genuine
structural information about how the interacting and noninteracting
hierarchies are related.

\section*{Concluding remarks}

This work proposes a formal reorganization of exact DFT into two parallel
ensemble variational hierarchies, an interacting ensemble theory and a
noninteracting ensemble theory, linked by the Kohn--Sham auxiliary
construction on their common \(N\)-representable density domain.

From this viewpoint, Lieb's ensemble formulation is the natural convex starting point for the interacting theory, while the analogous ensemble formulation of the exact noninteracting problem, with its own constrained-search and dual energy structure, provides the corresponding starting point on the noninteracting side. Along the state-class axis, the Levy--Lieb functional and the single-determinant noninteracting theory arise by restricting the admissible fibers to pure states and Slater determinants, respectively. These restricted formulations coincide with the corresponding ensemble functionals whenever an ensemble minimizer can be chosen within the restricted state class. Along the separate source-support axis, the Hohenberg--Kohn structure
expresses, on the corresponding \(v\)-representable subdomains, uniqueness of the supporting scalar potential modulo the additive-constant gauge within a fixed-particle-number sector. Exact Kohn--Sham theory is then the auxiliary construction that couples the two hierarchies on their common \(N\)-representable density domain; its familiar pure-state orbital realization is the corresponding state-class specialization, valid under \(N\)- and pure-state \(v\)-representability.

The same organization admits a compact optimization dictionary:
\(N\)-representability is primal feasibility, the Lieb relation is value
duality, \(v\)-representability is dual attainment, and the
Hohenberg--Kohn statement concerns uniqueness of the attained dual
potential modulo constants. The noninteracting hierarchy has the same
structure, while exact Kohn--Sham theory couples the interacting
density-space optimum to a compatible noninteracting dual realization.
Within this common language, piecewise linearity and the derivative
discontinuity are primal- and dual-side manifestations of nonsmooth
particle-number optimization, while Janak-type relations are envelope
derivatives of stationary orbital--occupation values. The same distinction
also separates statements about functional values from those requiring an
attained supporting potential, and exact density reproduction from a
general spectral interpretation of Kohn--Sham orbital energies.

The same perspective sharpens the status of the interaction remainder.
The primary interface quantity is \(\EHxc[\rho]=\Flieb[\rho]-\Ts[\rho]\). After the explicit
Hartree contribution is separated, \(\Exc\) records the remaining
difference between the interacting and noninteracting hierarchies and is
not merely an unknown term introduced for practical approximation.

\bibliography{ref}

\end{document}